\documentstyle[aps,preprint,tighten,floats,epsf,rotate,prl]{revtex}

\newbox\rotbox

\begin{document}
%\draft
\preprint{
\vbox{\noindent Submitted to {\it Physics Letters B} 
\hfill UW-PP-DOE/ER/40427-16-N95\\
\null \hfill TRI-PP-95-14\\
}} 
%%%%%%%%%%%%%%%%%%%%%%%%%%%%%%%%%%%%%%%%%%%%%%%%%%%%%%%%%%%%%%%%%%
 
\title{\LARGE Mesonic Width Effects on the Momentum Dependence of the
\\ $\rho-\omega$ Mixing Matrix Element}

\author{\sc M.J. Iqbal,$^{\rm a}$ Xuemin Jin,$^{\rm b}$ and 
Derek B. Leinweber$^{\rm b,c}$} 
\address{$^{\rm a}$Physics Department, University of British Columbia,
Vancouver, B.C. V6T 1Z1, Canada}
\address{$^{\rm b}$TRIUMF, 4004 Wesbrook Mall, Vancouver, B.C. V6T 2A3,
Canada} 
\address{$^{\rm c}$
Department of Physics, Box 351560, University of 
Washington, Seattle WA 98195, USA}
                                             
\maketitle

\begin{abstract}
It is shown, in a model independent way, that the large difference in
$\rho$ and $\omega$ widths gives rise to a new source of momentum
dependence for the $\rho-\omega$ mixing matrix element. The $q^2$
dependence arising due to the meson widths leads to a significant
alteration of the result obtained in the zero-width approximation
usually discussed in the literature. The origin of this strong
momentum dependence lies in the difference between the $\rho$ and
$\omega$ meson widths.
\end{abstract}
%
%\pacs{PACS number(s):11.30.-j,13.75.Cs,14.40.Cs}
\newpage
%%%%%%%%%%%%%%%%%%%%%%%%%%%%%%%%%%%%%%%%%%%%%%%%%%%%%%%%%%%%%%%%%%%%

Experimentally, the mixing of $\rho$ and $\omega$ mesons has been
observed in the G-parity forbidden decays of the $\omega$ meson,
$\omega \rightarrow \pi^{+} \pi^{-}$.  The generally accepted value
for the mixing matrix element is $\left< \rho | H_{\rm csb} | \omega
\right>$ = $-0.00452 \pm 0.00006 $ GeV$^{2}$\cite{coon1}.  This value
is extracted at $q^{2} \simeq m_{\omega} ^{2}$ and is often used to
generate the NN potential in a meson exchange approach
\cite{henley1,coon1}.  Even though the $\rho$ and $\omega$ mesons are
off-shell, it is common to find the use of the on-shell value.  Thus
there is an implicit assumption that the mixing matrix element is
$q^2$ independent.

Recently, it has been argued that the $\rho-\omega$ mixing matrix
element, extracted from the on-mass-shell mesons, should change
significantly off-shell\cite{goldman1,krein1,piekarewicz1,%
hatsuda1,maltman1,mitchell1,mitra1,connell1}. This off-shell behavior
significantly alters the contribution of $\rho-\omega$ mixing to the
CSB NN potential and its effects in $n-p$ scattering\cite{iqbal1}.
Theoretical calculations of the off-shell variations of the
$\rho-\omega$ mixing matrix element have used various models that
include mixing through $ q\overline q $ loops \cite{goldman1,krein1},
$N\overline N$ loops\cite{piekarewicz1}, a QCD sum rule calculation
\cite{hatsuda1,maltman1}, and a hybrid quark-meson coupling model
\cite{mitchell1,mitra1}.  In all of these calculations, the $\rho$ and
$\omega$ mesons are treated as stable particles and their decay widths
are neglected.

In this Letter, we shall show, in a model independent way, that the
large difference in $\rho$ and $\omega$ widths ($\Gamma_{\rho}$ =
151.5 MeV, $\Gamma_\omega$ = 8.4 MeV) gives rise to a new source of
momentum dependence for the $\rho-\omega$ mixing matrix element.  The
$q^2$ dependence arising due to the meson widths leads to a
significant alteration of the result obtained in the zero-width
approximation, typically discussed in the literature\cite{goldman1,%
krein1,piekarewicz1,hatsuda1,maltman1,mitchell1}. Thus in our view,
any discussion of the $q^2$ dependence of the $\rho-\omega$ mixing
matrix element that does not include the finite width effects of the
$\rho$ and $\omega$ mesons is incomplete.

Let us start from the $\rho$ and $\omega$ mixed
propagator\cite{hatsuda1}
\begin{equation}
\Pi^{\rho \omega}_{\mu\nu}(q)=i\int d^4x e^{iq\cdot x} \langle
0|T\rho_\mu (x)\omega_\nu(0)|0\rangle \equiv -\left(g_{\mu\nu}-{q_\mu
q_\nu\over q^2}\right)\Pi^{\rho \omega}(q^2)\ ,
\label{prop-def}
\end{equation}
where $\rho_\mu$ and $\omega_\nu$ are interpolating fields
representing the $\rho$ and $\omega$ mesons, respectively.
The analytic structure of the mixed propagator allows us to
write a dispersion relation of the form,
\begin{equation}
{\rm Re\ } \Pi^{\rho \omega}(q^2) = \frac{P}{\pi} \,
\int_{4 m_{\pi}^2}^{\infty}  \frac{{\rm Im}\, \Pi^{\rho \omega}(s)
}{(s-q^2)} ds \, ,
\label{dis-gen}
\end{equation}
which is valid to leading order in quark masses~\cite{hatsuda1}.

The mixing matrix 
element $\theta(q^2)$ has the following definition~\cite{hatsuda1}
\begin{equation}
{\rm Re\ } \Pi^{\rho \omega}(q^2) \equiv  
\frac{\theta(q^2)}{(q^2-m_{\rho}^2)
(q^2 - m_{\omega}^2 )} \, .
\label{m-def-ha}
\end{equation}
Traditionally $\theta(q^2)$ is regarded as mixing between
$\rho$ and $\omega$ ground states. However, Eq.~(\ref{dis-gen})
indicates that $\theta(q^2)$ must also include physics of excited
states. Hence to make contact with the traditional phenomenology,
one must select $\rho$ and $\omega$ interpolating fields which
have maximal overlap with the ground state mesons. 

In fact,
it has been argued recently that there is no unique choice of 
the interpolating fields $\rho_\mu$ and $\omega_\nu$ and 
the mixed propagator depends on the choice of interpolating 
fields\cite{cohen1,maltman2}. Although this is certainly  the 
case, there is, however, an obvious and physically reasonable 
choice which is consistent with standard traditional phenomenology. 
Namely, one should select interpolating fields with maximal 
overlap with ground states. One could introduce wave functions 
to smear out the relative separation of the quark field operators 
in the interpolating fields to improve the overlap with ground states
as done in lattice calculations. However, such a nonlocal
approach is not gauge invariant. Alternatively one might 
consider interpolating fields involving gluon field strength
operators or derivatives. However, such interpolating fields 
are of higher dimension and have an increased overlap with 
excited states relative to the ground states.
Hence the preferred interpolating fields most consistent with the
traditional phenomenology are local and of minimal dimension.
The standard vector currents satisfy these criteria.

\begin{figure}[t]
\begin{center}
\epsfysize=11.6truecm
\leavevmode
\setbox\rotbox=\vbox{\epsfbox{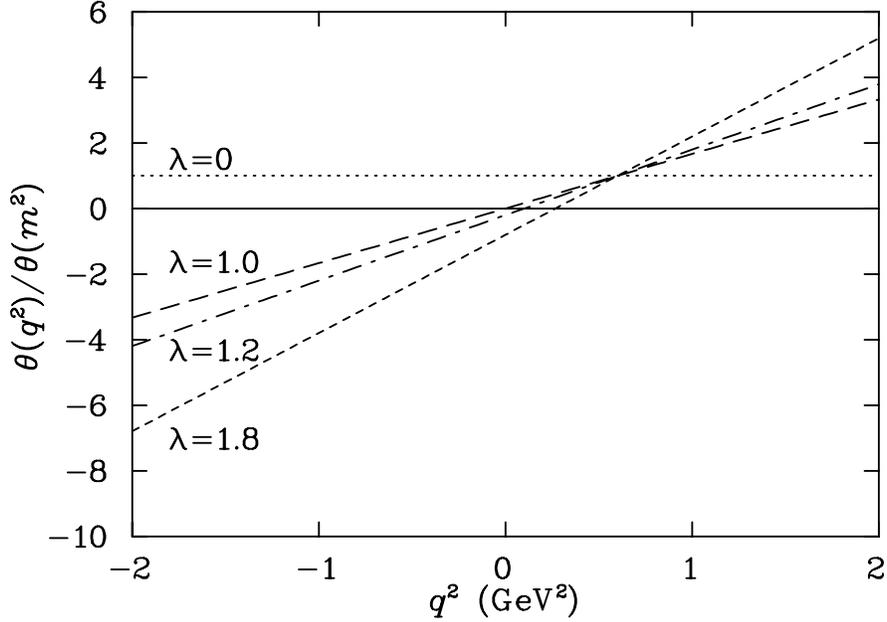}}\rotl\rotbox
\end{center}
\caption{The normalized $\rho-\omega$ mixing amplitude
$\theta(q^2)/\theta(m^2)$ plotted as a function of $q^2$
for various values of $\lambda$.}
\label{fig-1}
\end{figure}

To illustrate the effect of including the finite widths of the 
mesons on the mixing matrix element, it is sufficient to saturate 
the imaginary part of the mixed propagator on the
right hand side of (\ref{dis-gen}) with $\rho$ and $\omega$ mesons
alone. We begin with the zero-width approximation,
\begin{equation}
{\rm Im}\, \Pi^{\rho \omega} (s) =  \pi \, F_{\rho} \, 
\delta(s-m_{\rho}^2) -
\pi \, F_{\omega} \, \delta (s-m_{\omega}^2) \, .
\label{saturation}
\end{equation}
Here $F_{\rho}$ and $ F_{\omega}$ denote the coupling strengths of the
interpolating fields to the physical meson states.  While a more
sophisticated treatment of the spectral density or dispersion relation
might be desirable, such complications unnecessarily obscure the
qualitative physics we are emphasizing here.  Equation~(\ref{dis-gen})
now takes the form
\begin{equation}
{\rm Re\ }  \Pi^{\rho \omega}(q^2) =
\frac{-F_{\rho} (q^2-m_{\omega}^2) + F_{\omega} (q^2-m_{\rho}^2)}
{(q^2-m_{\rho}^2)(q^2-m_{\omega}^2)}\, ,
\label{tot-0}
\end{equation}
which implies
\begin{equation}
\theta (q^2)  =  \frac{1}{2} \, (F_{\rho} + F_{\omega}) \, \delta m^2
- (F_{\rho} - F_{\omega}) (q^2 - m^2)\ .
\label{theta0}
\end{equation}
Here we have introduced the notation 
$\delta m^2 \equiv m_{\omega}^2 - m_{\rho}^2$ and
$m^2 \equiv (m_{\omega}^2 + m_{\rho}^2)/2$.
One may then express $\theta(q^2)$ in terms of
$\theta(m^2)=(F_{\rho} + F_{\omega}) \, \delta m^2/2$ as
\begin{equation}
\theta(q^2) = \theta(m^2) \left [ 1 + \lambda \; (\frac{q^2}{m^2} -1 )
\right ] \, ,
\label{theta-hats}
\end{equation}
where $\lambda \equiv  - m^2(F_{\rho} - 
F_{\omega})/ \theta (m^2)$,
and $\theta(m^2)$ is the on-shell $\rho-\omega$ mixing matrix element.
Note that the $q^2$ dependence of the
mixing matrix element arises due to the second term of
Eq.~(\ref{theta0}). If $F_{\rho} = F_{\omega}$, implying $\lambda =
0$, then there is no $q^2$ dependence.  Also note that the sign of
$\theta (q^2)$ changes at $q^2 = m^2 (\lambda -1) /{\lambda}$.
Hatsuda {\it et al.}\cite{hatsuda1} have extracted the parameter 
$\lambda$ from a QCD sum rule analysis and quoted the values 
$1.43 \leq \lambda \leq 1.85$. While we have a few reservations 
regarding their analysis, we will use their 
estimate of $\lambda$ to illustrate the effects of mesonic widths 
and their relative importance in determining the $q^2$ dependence 
of $\rho-\omega$ mixing.  We will also consider other values to 
more clearly illustrate our findings. In Fig.~\ref{fig-1} we have 
plotted the ratio $\theta (q^2) / \theta(m^2)$ for various values 
of $\lambda$.

In the above discussion, we have assumed zero-width sharp poles 
for the mesons. In nature, however, both $\rho$ and $\omega$
are resonances with finite widths. To include these finite widths, 
we replace the $\delta$-functions in Eq.~(\ref{saturation}) by a 
normalized Breit-Wigner form
\begin{equation}
\frac{1}{\pi} \frac{m \, \Gamma}
  {(q^2-\overline m^2)^2  + m^2 \, \Gamma ^2} \, ,
\end{equation}
where $\overline m^2 = (m^2 - \Gamma ^2 / 4)$, and $\Gamma$ is the
half width of the meson. This form has a smooth limit to the
$\delta$-function. In the absence of a solution to QCD, the precise
shape of the spectral density is unknown. As such, we consider a sum
of the $\rho$ and $\omega$ resonant peaks with finite widths, which
should provide a physically reasonable model for the spectral
density. With the following integral definition
\begin{equation}
I_{\rho , \omega} (q^2) \equiv \frac{m_{\rho , \omega} \Gamma_{\rho ,
\omega}} {\pi}
P \int_{4m_{\pi}^2}^{\infty}
\frac{ds}{(s-q^2) \left [ (s-\overline{m}_{\rho, \omega}^2)^2
+m_{\rho , \omega}^2 \, \Gamma_{\rho , \omega}^2) \right ] } \, ,
\end{equation}
we can write the analogue of Eq.~(\ref{tot-0}) for 
${\rm Re\ } \Pi^{\rho \omega}$ in a compact
form\cite{cutoff}
\begin{equation}
{\rm Re\ }  \Pi^{\rho \omega}(q^2) =  F_{\rho} \, I_{\rho}(q^2) - 
F_{\omega} \, I_{\omega}(q^2) \, .
\label{real-f}
\end{equation}
This expression reduces to the zero-width result of (\ref{tot-0}) in
the limit of the meson widths going to zero.

Let us first adopt the simple zero-width form of Eq.~(\ref{m-def-ha})
to define the mixing matrix element $\theta_1(q^2)$
\begin{equation}
{\rm Re\ }  \Pi^{\rho \omega}(q^2) =  \frac{\theta_1(q^2)}{ (q^2 -
m_{\rho} ^2) (q^2 - m_{\omega} ^2 ) } \, ,
\label{zerowidth}
\end{equation}
where ${\rm Re\ }  \Pi^{\rho \omega}(q^2)$ is now given by
Eq.~(\ref{real-f}). This allows an examination of how the
inclusion of finite meson widths affects the traditionally defined
$\theta(q^2)$. With the definition
$G_{0} (q^2)  \equiv  (q^2 - m_{\rho} ^2) (q^2 - m_{\omega} ^2 )$
we have
\begin{equation}
\theta_{1} (q^2) = G_{0} (q^2) \left [ F_{\rho} \, I_{\rho}(q^2) - 
F_{\omega} \, I_{\omega}(q^2) \right ] \, .
\label{theta1-pre}
\end{equation}

\begin{figure}[b]
\begin{center}
\epsfysize=11.6truecm
\leavevmode
\setbox\rotbox=\vbox{\epsfbox{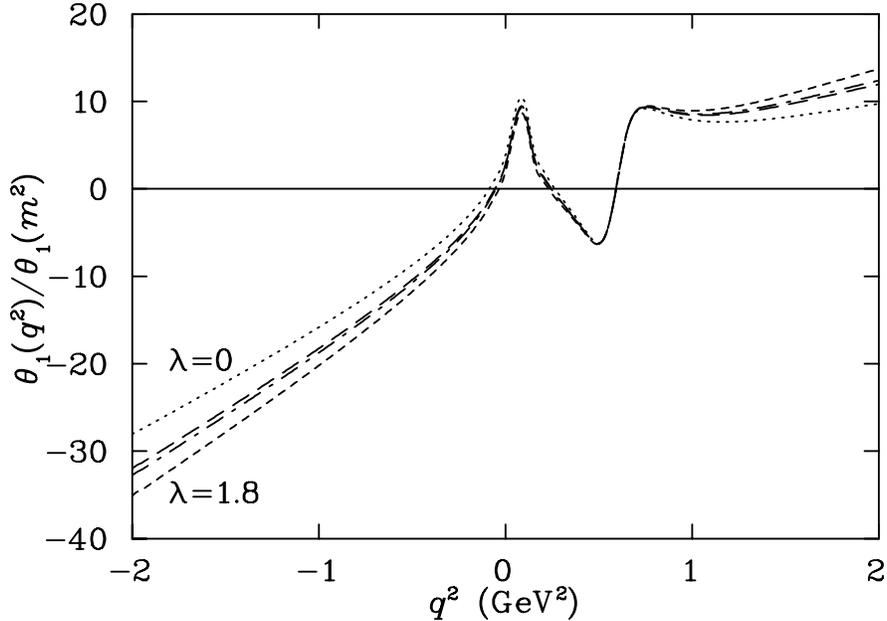}}\rotl\rotbox
\end{center}
\caption{$\theta_{1}(q^2)/\theta_1(m^2)$ as a function of $q^2$ for
the same values of $\lambda$ as used in Fig.~\protect{\ref{fig-1}},
with $\Gamma_{\rho} = 151.5$ MeV and $\Gamma_{\omega} = 8.4$ MeV.}
\label{fig-2}
\end{figure}

One can rewrite $\theta_1(q^2)$ in terms of $\theta_1(m^2)$ as
\begin{equation}
\theta_{1} (q^2)  =  \theta_1 (m^2) \Biggl \{
\frac{G_{0}(q^2)}{G_{0}(m^2)} 
\, \frac{I_{\rho} (q^2) - I_{\omega} (q^2) }
{I_{\rho} (m^2) - I_{\omega} (m^2) }
+ \lambda \;
\frac{G_{0} (q^2)}{m^2} \,
\frac{ I_{\rho} (q^2) I_\omega (m^2) - I_{\omega} (q^2) I_\rho(m^2)}
{I_{\rho} (m^2) - I_{\omega} (m^2)} 
\Biggr \} \, ,
\label{theta2f}
\end{equation}
where $\lambda \equiv - m^2 \, (F_{\rho}-F_{\omega}) / \theta_1(m^2)$,
with $\theta_1(m^2)$ the on-shell value.  Now both terms in braces are
$q^2$ dependent.  This arises because the widths for $\rho$ and
$\omega$ mesons are very different, and this leads to a different
$q^2$ dependence in the integrals $I_{\rho }$ and $I_{\omega}$.  In
Fig.~\ref{fig-2} we have plotted $\theta_{1} (q^2) / \theta_1(m^2)$ as
a function of $q^2$ for various values of $\lambda$, with the physical
values for the meson widths.  A comparison with Fig.~\ref{fig-1}
reveals significant alteration of the $q^2$ dependence for all values
of $\lambda$.

\begin{figure}[t]
\begin{center}
\epsfysize=11.6truecm
\leavevmode
\setbox\rotbox=\vbox{\epsfbox{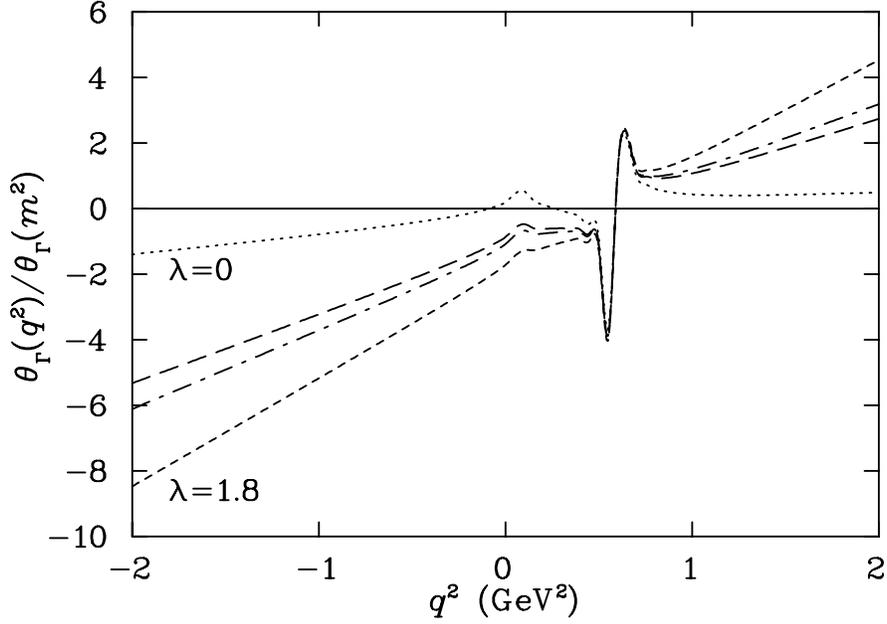}}\rotl\rotbox
\end{center}
\caption{$\theta_{\Gamma}(q^2)/\theta_{\Gamma}(m^2)$ as a function of
$q^2$ for values of $\lambda=0,\, 1.0,\, 1.2,$ and $1.8$, with
$\Gamma_{\rho} = 151.5$ MeV and $\Gamma_{\omega} = 8.4$ MeV.}
\label{fig-3}
\end{figure}

   To be fully consistent, meson widths should also be included in 
extracting the mixing matrix element from $\Pi^{\rho\omega}$
and we relate the mixing matrix element to $\Pi^{\rho\omega}(q^2)$
by
\begin{equation}
\Pi^{\rho \omega}(q^2) \equiv \frac{\theta_{\Gamma} (q^2)}{ \left[q^2
- (m_{\rho} -i \frac{\Gamma_{\rho}}{2}) ^2 \right]\left[q^2 -
(m_{\omega} -i \frac{\Gamma_{\omega}}{2}) ^2 \right]} \, ,
\label{full}
\end{equation}
where $\Gamma_{\rho}$ and $\Gamma_{\omega}$ are $q^2$-dependent.  In
particular, below threshold ($q^2=4m_\pi^2$), $\Gamma_{\rho,\omega}=0$
and $\Pi^{\rho \omega}(q^2 <4m_\pi^2)$ is real as it should be.  Above
threshold, we take $\Gamma_{\rho}$ and $\Gamma_{\omega}$ to be
constant for simplicity. Here we have introduced the notation
$\theta_{\Gamma}$ to indicate the fully consistent definition of the
mixing matrix element.  The real part $\Pi^{\rho \omega}$ is now given
by
\begin{equation}
{\rm Re\ } \Pi^{\rho \omega}(q^2) = \theta_{\Gamma}(q^2) \frac
{(q^2 - \overline m_\rho^2)(q^2 - \overline m_\omega^2) -
m_{\rho} \, m_{\omega} \,
\Gamma_{\rho} \, \Gamma_{\omega} }
{ \left[(q^2 - \overline m_\rho^2 )^2 + m_{\rho}^2 \Gamma_{\rho}^2
\right] 
 \left [ (q^2 - \overline m_\omega^2 )^2 + m_{\omega}^2
\Gamma_{\omega}^2 \right ] } 
\, .
\label{full-r}
\end{equation}
Recall, $\overline m^2_{\rho,\omega} = (m^2_{\rho,\omega} 
- \Gamma^2_{\rho,\omega} / 4)$. As before, we can
write $\theta_{\Gamma}(q^2)$ in terms of $\theta_{\Gamma}(m^2)$ as
\begin{equation}
\theta_{\Gamma} (q^2)  =  \theta_\Gamma (m^2) \Biggl \{ 
\frac{ G(q^2)}{G(m^2)}
\, \frac{ I_{\rho} (q^2) - I_{\omega} (q^2) }
{ I_{\rho} (m^2) - I_{\omega} (m^2) }
+\lambda \;
\frac{G(q^2)}{m^2}
\frac{ I_{\rho} (q^2)I_\omega (m^2) 
- I_{\omega} (q^2)I_\rho (m^2) }
{I_{\rho} (q^2) - I_{\omega} (q^2)}
\Biggr \} \, ,
\label{theta3f}
\end{equation}
where $G(q^2)$ is defined as
\begin{equation}
G(q^2) =  \frac
{ \left[(q^2 - \overline m_{\rho} ^2 )^2 + m_{\rho}^2 \,
\Gamma_{\rho}^2 \right] 
\left [ (q^2 - \overline m_{\omega} ^2 )^2 + m_{\omega}^2 \,
\Gamma_{\omega}^2 \right ] } 
{ (q^2 - \overline m_{\rho} ^2)(q^2 - \overline{m}_{\omega}^2) -
m_{\rho} \, m_{\omega} \, 
\Gamma_{\rho} \, \Gamma_{\omega} } \, ,
\end{equation}
and $\lambda\equiv -m^2 (F_\rho-F_\omega)/\theta_{\Gamma}(m^2)$, with
$\theta_{\Gamma}(m^2)$ the on-shell value. We note that the relation
of (\ref{full}) provides better contact with the experimentally
extracted on-shell value than that given in (\ref{m-def-ha}).  A plot
of $\theta_{\Gamma}(q^2)$ for various values of $\lambda$ is shown in
Fig.~\ref{fig-3}.  The $q^2$ dependence of $\theta_{\Gamma}(q^2)$ is
softer than that for $\theta_{1}(q^2)$.  We observe that the
expression for $\theta_\Gamma(q^2)$ in (\ref{theta3f}) is $q^2$
dependent regardless of the value taken for $\lambda$. This is in
accord with the arguments of Ref.~\cite{maltman2} based on unitarity
and analyticity. 

   Recently, arguments have been made for the vanishing of
$\Pi^{\rho\omega}(q^2)$ at $q^2$. The arguments are quite general and
apply to any effective field theory in which there is no explicit
mass-mixing term in the bare Lagrangian. Unfortunately, we are unable
to comment on the viability of such models without knowledge of the
value for $\lambda$. A fundamentally based QCD determination of
$\lambda$ is forthcoming\cite{iqbal2}.

\begin{figure}[t]
\begin{center}
\epsfysize=11.6truecm
\leavevmode
\setbox\rotbox=\vbox{\epsfbox{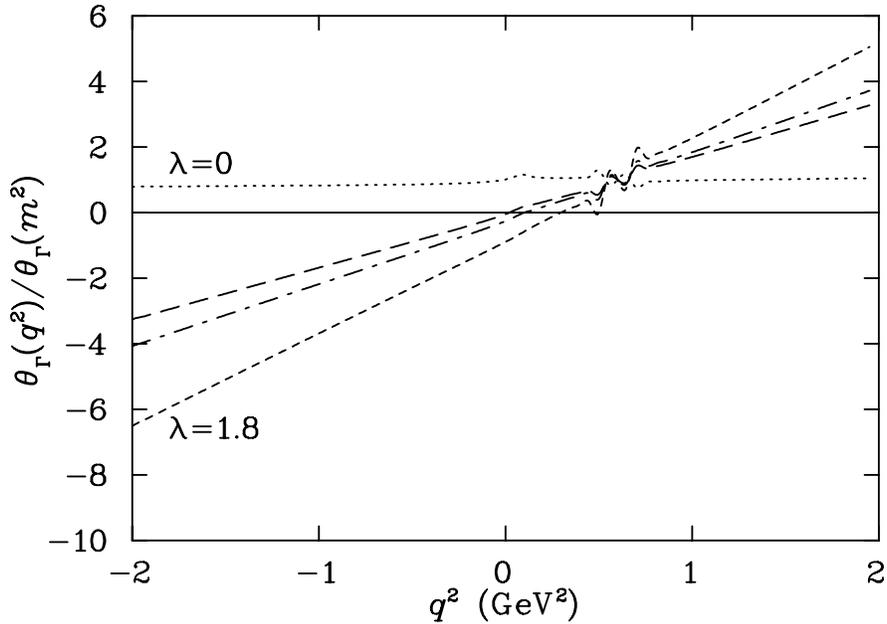}}\rotl\rotbox
\end{center}
\caption{$\theta_{\Gamma}(q^2)/\theta_{\Gamma}(m^2)$ as a function of
$q^2$, with $\Gamma_{\rho}$= $\Gamma_{\omega} = 100$ MeV. The values
of $\lambda$ are the same as those used in previous figures.}
\label{fig-4}
\end{figure}

For comparison, we present in Fig.~\ref{fig-4} the ratio
$\theta_{\Gamma}(q^2)/\theta_{\Gamma}(m^2)$ for the same values of
$\lambda$ as in Fig.~\ref{fig-3}, but with $\Gamma_{\rho}$ =
$\Gamma_{\omega}$ = 100 MeV.  The $q^2$ dependence in this case is
similar to that for the zero-width case.

\begin{figure}[t]
\begin{center}
\epsfysize=11.6truecm
\leavevmode
\setbox\rotbox=\vbox{\epsfbox{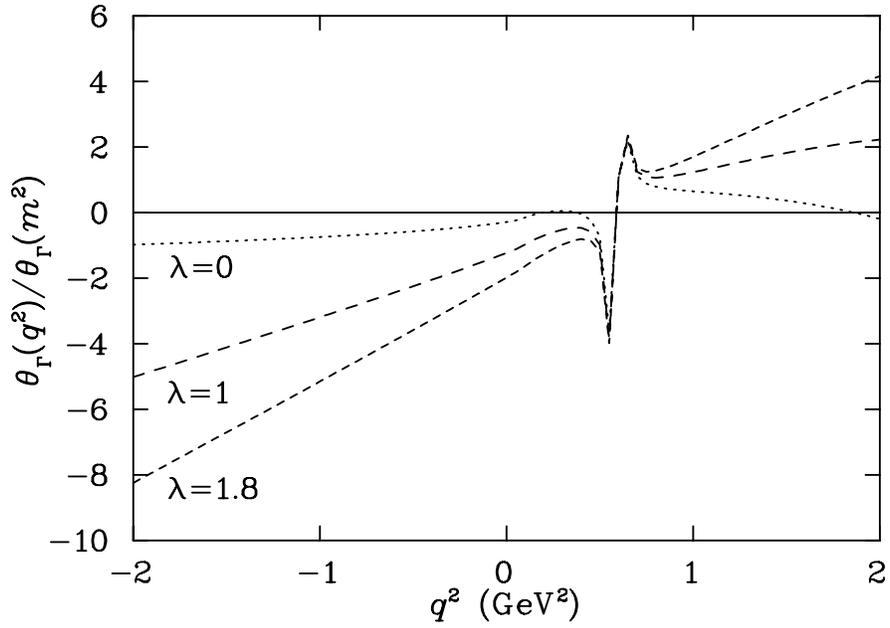}}\rotl\rotbox
\end{center}
\caption{$\theta_{\Gamma}(q^2)/\theta_{\Gamma}(m^2)$ as a function of
$q^2$ with an energy dependent width for the $\rho$ meson as given in
(\protect\ref{ed-width}).}
\label{fig-5}
\end{figure}

The effects of including an energy dependent $\rho$-meson
width\cite{energyd} on the right-hand side of the dispersion relation
of~(\ref{dis-gen}) are illustrated in Fig.~\ref{fig-5}.  Here we
adopt the standard form\cite{energyd}
\begin{equation}
\Gamma_{\rho}(s)=\Gamma_{\rho}(m_\rho^2)\, {m_\rho\over \sqrt{s}}
\left( {s-4 m_\pi^2\over m_\rho^2-4 m_\pi^2}\right)^{3/2},\quad s\ge
4m_\pi^2\ ,
\label{ed-width}
\end{equation}
normalized to the previous value at $s = m_\rho^2$.  The predominant
effect is a smoothing of the curves in the threshold regime. The
oscillation seen in Fig.~\ref{fig-3} is an artifact of the abrupt
onset of the $\rho$-meson width at the threshold $4 \, m_\pi^2$ as
discussed following Eq.~(\ref{full}).  The energy dependence of the
$\rho$ width does not alter our general conclusions.

In summary, we have shown in a model independent way that the
inclusion of $\rho$ and $\omega$ widths significantly alters the $q^2$
dependence of the $\rho-\omega$ mixing matrix element and hence of the
mixed meson propagator. This behavior arises from the fact that the
widths of $\rho$ and $\omega$ are different.  Any model calculation
addressing the $q^2$ dependence of the $\rho-\omega$ mixing matrix
element that does not include meson width effects is incomplete.

\newpage
\acknowledgements

M.J.I acknowledges useful discussions with David Axen regarding the 
energy dependence of the $\rho$ meson width.
This work was supported by the Natural Sciences and Engineering
Research Council of Canada and the US Department of Energy
under grant DE-FG06-88ER40427.
%\newpage


\begin{references}
\bibitem{cutoff}The integrals $I_\rho$ and $I_\omega$ are cut off
at $s=3\,\text{GeV}^2$. This value is selected as the integral
values are stable and the physics of excited resonances is expected 
to alter the spectral density.
\end{references}
\end{document}